\documentclass[aps,prl,floats,twocolumn,superscriptaddress]{revtex4}
\usepackage{graphicx, amsfonts}
\usepackage[latin1]{inputenc}
\usepackage[T1]{fontenc}

\begin{document}

\title{Sub-Terahertz Monochromatic Transduction with Semiconductor Acoustic Nanodevices}
\author{A. Huynh}
\author{B. Perrin}
\affiliation{Institut des Nanosciences de Paris, Universit\'e Paris 6, CNRS, 140 rue de Lourmel 75015 Paris, France}
\author{N. D. Lanzillotti-Kimura}
\affiliation{Institut des Nanosciences de Paris, Universit\'e Paris 6, CNRS, 140 rue de Lourmel 75015 Paris, France}
\affiliation{Centro At\'omico Bariloche and Instituto Balseiro, C.N.E.A., 8400 S. C. de
Bariloche, Argentina}
\author{B. Jusserand}
\affiliation{Institut des Nanosciences de Paris, Universit\'e Paris 6, CNRS, 140 rue de Lourmel 75015 Paris, France}
\author{A. Fainstein}
\affiliation{Centro At\'omico Bariloche and Instituto Balseiro, C.N.E.A., 8400 S. C. de
Bariloche, Argentina}
\author{A. Lema\^itre}
\affiliation{Laboratoire de Photonique et de Nanostructures, CNRS, Route de Nozay, 91460
Marcoussis, France}

\begin{abstract}
We demonstrate semiconductor superlattices or nanocavities as narrow band acoustic transducers in the sub-terahertz range. Using picosecond ultrasonics experiments in the transmission geometry with pump and probe incident on opposite sides of the thick substrate, phonon generation and detection processes are fully decoupled. Generating with the semiconductor device and probing on the metal, we show that both superlattices and nanocavities generate spectrally narrow wavepackets of coherent phonons with frequencies in the vicinity of the zone center and time durations in the nanosecond range, qualitatively different from picosecond broadband pulses usually involved in picosecond acoustics with metal generators. Generating in the metal and probing on the nanoacoustic device, we furthermore evidence that both nanostructured semiconductor devices may be used as very sensitive and spectrally selective detectors.

\end{abstract}

\date{\today}
\maketitle


The availability of efficient monochromatic phonons
transducers in the THz range would give access to spectroscopic
studies of vibrational properties in amorphous
and quasi-crystalline materials, in nanostructures or in
any system exhibiting inhomogeneities at a nanometer
scale. These waves could also be used for high resolution
acoustic microscopy and to drive the optical and electronic
properties of devices at the picosecond time scale. Piezoelectric transduction is
very efficient up to a few GHz but only special geometries
can work beyond 100 GHz\cite{jacobsen68}. Propagation of phonons in
the THz range by heat pulses techniques has been extensively
studied since the pioneering work of Von Gutfeld\cite{neth64} but these phonons are incoherent and very broadband.Tunneling junctions\cite{junctiontunnel3,junctiontunnel4} provide monochromatic but still incoherent sources. 
Later, coherent acoustic wave generation and detection by an optical pump
probe technique and a femtosecond laser source proved
to be very successful\cite{thomsenMaris}; using a metallic film as a transducer,
generation and detection of very short acoustic
pulses with a broad frequency spectrum extending up to a few
hundreds GHz can be performed. Similar results have
been demonstrated more recently but with semiconductor
single quantum wells (QW) instead of metallic layers as phonon
generators or detectors\cite{QW}. QW transducers present the advantage
to behave as internal probes with an almost arbitrary
location inside the studied devices. However they
remain broadband transducers with a high frequency cut-off 
limited by the QW thickness.  
Regularly stacking several QWs and then building superlattices (SL) should result in an enhancement 
of the transduction efficiency. More important, SL offer a unique access to high frequency monochromatic acoustic phonon optical transduction. Indeed, high frequency folded modes are strongly coupled
to light in SL and the THz range can be reached
with typical periods of a few nanometers, easily controlled
by modern deposition techniques. Moreover, the periodic modulation
of the elastic properties leads to the opening of energy
gaps in the Brillouin zone center and boundaries. Light scattering
has provided a detailed understanding on the SL vibrations
at thermal equilibrium\cite{cardonaJusserand} while ultrafast optics
experiments also showed that high frequency modes can
be coherently generated and controlled\cite{grahnMaris,chenmaris,perrinbonello,Mizoguchi2colors,Pu2,Bartels}. 
The initial proposal of SL as potentially ideal monochromatic optical transducers
for coherent phonons\cite{basseras} has however received up
to now very limited experimental demonstration. Propagation
through a substrate of acoustic pulses optically excited in SLs was first
reported using an incoherent detection with a superconducting
bolometer deposited on the other side of the sample\cite{kent1}. A subsequent experiment using
a second SL as an acoustic filter gave support to the
coherence of the emitted phonons\cite{kent2} though leaving the
spectral width of the generated phonons unknown. 

In this letter we give a direct evidence of the transduction 
of coherent monochromatic phonons using SL or recently introduced acoustic
nanocavities\cite{trigocavity,huynhcavity}, by detecting these phonons with a pump probe
technique after they propagate through a thick
substrate. Indeed, in contrast with classical pump and
probe experiments where both the generation and the detection are performed on the device surface, here the generation process takes place on one side of
the sample in the device whereas the detection is done on the backside
with a broadband metallic transducer. In opposition to metallic transducers, the generated phonons can not be described as a broadband pulse with a duration in the picosecond range but as quasi-monochromatic coherent oscillations with duration in the nanosecond range. Alternatively, the metallic layer is also
used as a coherent broadband generator to demonstrate that
semiconductor multilayers are very sensitive and
frequency selective phonon detectors. In these experiments, the propagation
of the generated acoustic signal over a large distance
allows the decoupling of generation and detection processes. 
Moreover, during propagation, nonlinear distortion of the initial acoustic signal is likely to occur at high displacement amplitudes, which could give rise to acoustic solitons\cite{marissoliton,peronnesoliton} and high
frequencies conversion. Sound attenuation is made negligible across the thick GaAs substrate (330 to 380$\mu $m) by cooling down samples to 15K. Finally we compare the transduction performance of the standard SL
and the nanocavity devices. 
We show
that the latter device provides an unsurpassed ultimate
line width associated with confined undispersive cavity
phonons while the overall behavior of the two devices
remains qualitatively similar.

Owing to the limitations of the aluminium transducer, 
we choose to work on semiconductor structures grown on double side polished [001] GaAs substrate, 
with a zone center first gap located at 100GHz or 200GHz. The first ones contain 35nm/15nm GaAs/AlAs SL, resulting in opening
gaps every 50GHz, whereas for 200GHz  SL, layers are twice thinner and gaps open every 100GHz. 
Sample M1 (resp. M2) is a 20-period (resp. 40) SL designed at 100GHz (resp. 200GHz). SL acting as an acoustic Bragg mirror, surface localized
modes can be confined in superficial layers of the mirror
since the interface with air is a perfect reflector\cite{grahnMaris, perrinbonello} but
an acoustic nanocavity can also be obtained if a GaAs layer is
sandwiched between two such mirrors. 
We present results on cavities containing 10-period Bragg mirrors : sample C1 (resp. C2) with an acoustic confined mode at 100GHz (resp. 200GHz) and a cavity thickness of 23nm (resp. 37nm). 
For all samples, the light absorption threshold is about 810nm at low temperature.

When a laser pump beam interacts with such nanostructures, an acoustic strain modulated at the superlattice period is
created, and generation of very high frequency coherent
phonons is expected in a narrow frequency range. To demonstrate this generation, we performed a first set of
experiments (named afterwards configuration 1, cf. inset
in fig.~1) where the pump pulse is focused directly
on the device, generating acoustic phonons which propagate
through the whole substrate. They are detected by
a time delayed probe pulse reflected by a 30nm thick aluminum layer
deposited on the other side of the sample. 
\begin{figure}[tbp]
\includegraphics[angle=-90,width=0.45\textwidth]{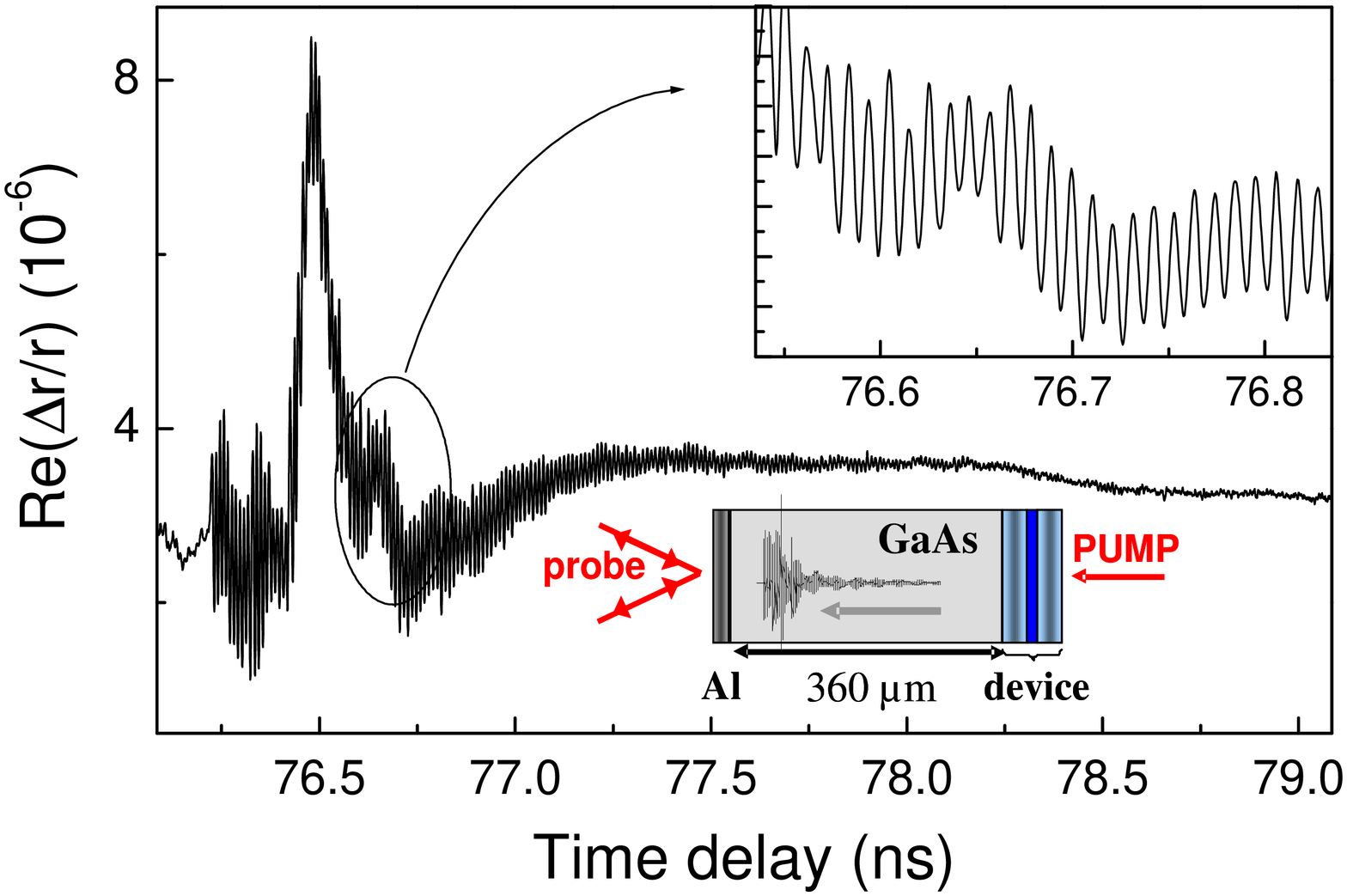} 
\caption{Re($\Delta r/r$) on sample M1 as a function of the pump-probe delay. The insets show an expanded region and a scheme of the picosecond ultrasonics measurement (configuration 1), where a calculated temporal trace of the strain generated by the device is drawn. }
\label{fig1}
\end{figure}
We used a femtosecond mode-locked
Ti:Sapphire laser providing 130fs pulses at 750nm, with a 80MHz  
repetition rate. At this wavelength, the absorption length in the device $1/\alpha$  is comparable to the multilayer thickness. The transient changes of the complex optical
reflectivity $\Delta r/r$ are measured by a Sagnac interferometer\cite{duquesneperrin}. The pump is modulated at 1MHz and the probe is detected by a lock-in amplifier. Both beams are focused
onto 60$\mu $m spots with a typical energy of a
few nJ/pulse for the pump. 

\begin{figure}[b]
\includegraphics[angle=-90,width=0.47\textwidth]{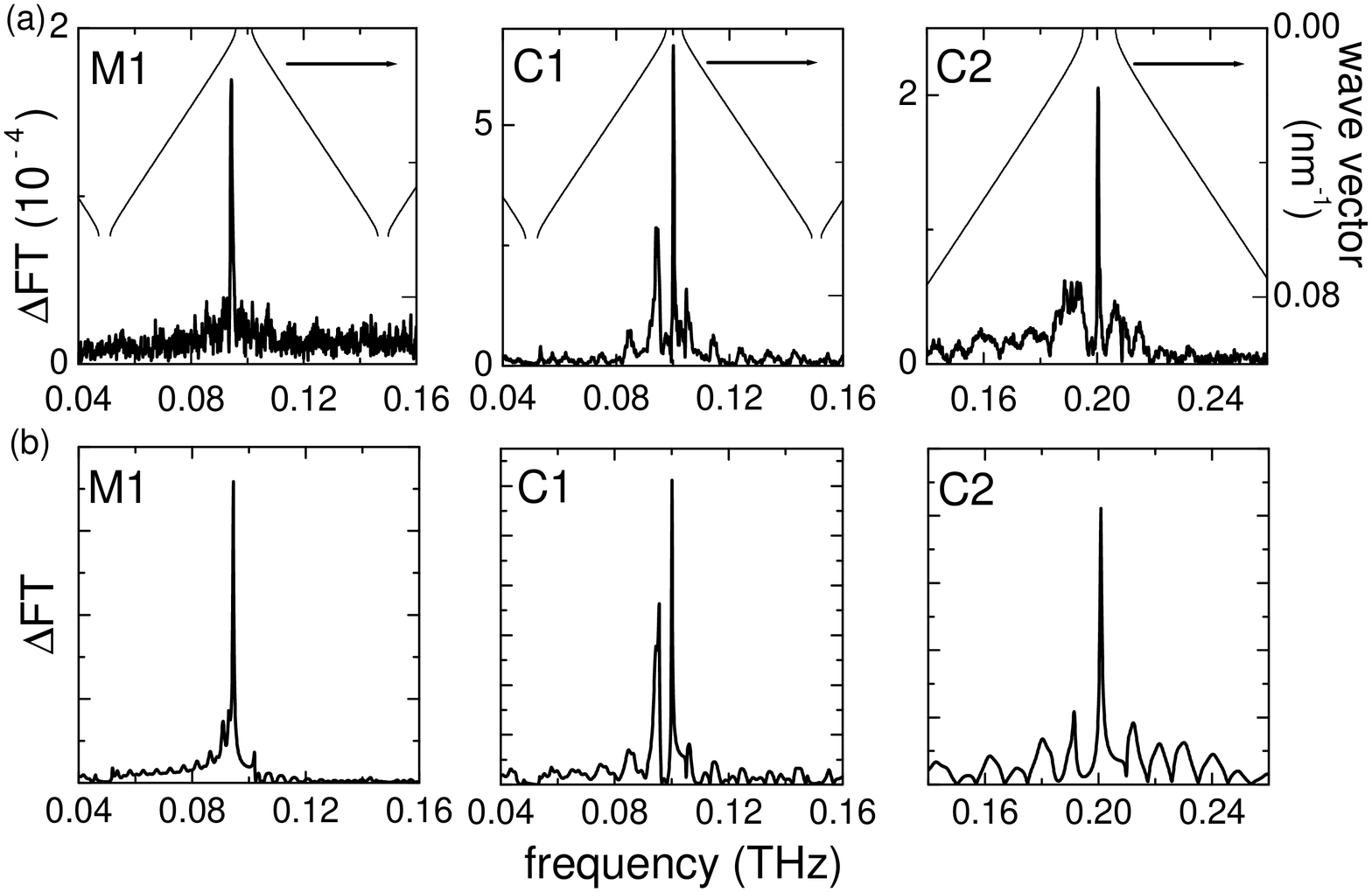} \vspace*{0cm}
\caption{Configuration 1. (a): Fourier transform amplitudes of the time derivative of Im($\Delta r/r$), compared to the folded phonon dispersion.  The pump power is 5nJ/pulse. (b): Corresponding calculations.}
\label{fig2}
\end{figure}
A typical signal (obtained on sample M1) corresponding to the real 
part of $\Delta r/r$ is shown in fig.~1. Due to
the very thick substrate, no direct electronic or thermal
contributions appear but a standard low frequency acoustic pulse. Its width amounts to 140ps in good agreement with the expected value $\alpha v^{-1}$ where $v$ is the averaged velocity in the device (5nm/ps). 
The maximum of the signal gives the arrival time of the acoustic pulse (76.5ns) taking into account acoustic reflections on the free surfaces of both aluminum film and device. Moreover, high frequency oscillations extending over a few nanoseconds with an amplitude of $10^{-6}$ are
clearly visible. They start 200ps before the signal maximum, a delay corresponding to the transit time through the device: the first oscillations are due to the acoustic
signal generated by the device deepest layers \textit{i.e.} the last ones before the substrate. 
The figure~2a shows the Fourier transform amplitude of the time derivative ($\Delta $FT) of the signals over a 3ns time window to ensure a sufficient frequency resolution (0.3GHz). Note that both real and imaginary part of $\Delta r/r$ give similar spectral information. The acoustic resonances of a mirror made of
$m$ periods $d$ occur at energies corresponding to wave vectors $q=n\pi /md$, $n\in \mathbb{N}$ but $\frac{n}{m} \notin \mathbb{N}$ in the extended Brillouin zone of the related infinite structure. Moreover, for a weakly absorbing structure, an
enhancement of the generation process is mostly expected near the zone center of the reduced Brillouin zone. 
Thus the lowest energy large peaks in the Fourier spectrum are expected
at $qd=2\pi \pm \pi /m$, among
smaller resonances. From symmetry considerations applied
to the lowest zone center gap, only the mode below lower gap edge should be excited in our stuctures;
it will be referred to the forward scattering (FS) mode as the same mode is Raman active in forward scattering configuration. The comparison of the experimental spectra with the dispersion curve (see fig.~2a) shows that the excited peak corresponds to this mode. As $m =20$, it is indeed slightly downshifted relatively to $q=0$ mode. 
Let us now consider sample C1. The FS mode can also be
observed, but at a larger distance from the zone center since
acoustic mirrors of the cavity have only $m = 10$ periods. 
The most remarkable fact, however, is the presence of a very narrow
cavity mode in the middle of the first zone center gap. 
\begin{figure}[b]
\includegraphics[angle=-90,width=0.25\textwidth]{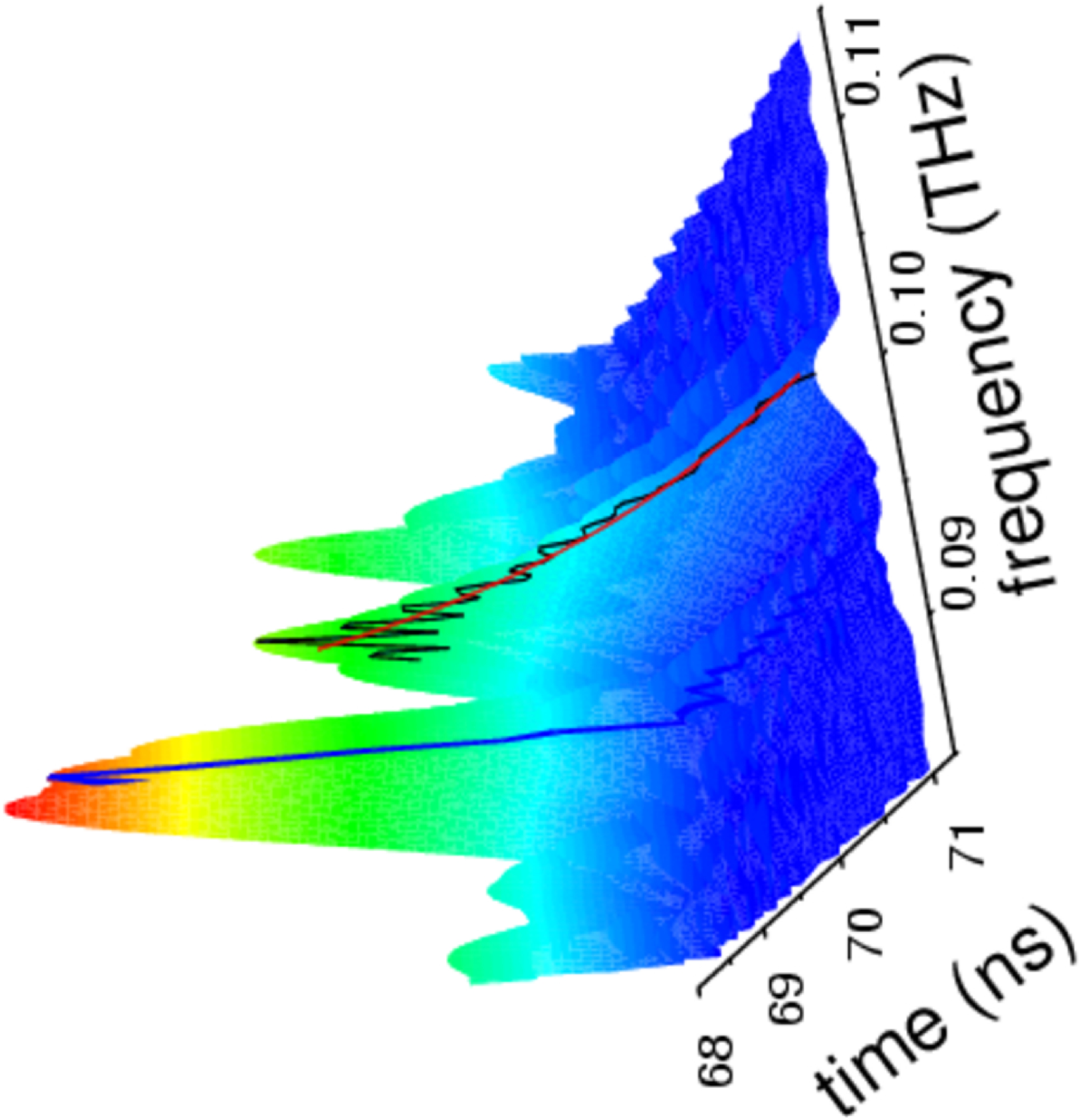} \includegraphics[angle=-90,width=0.2\textwidth]{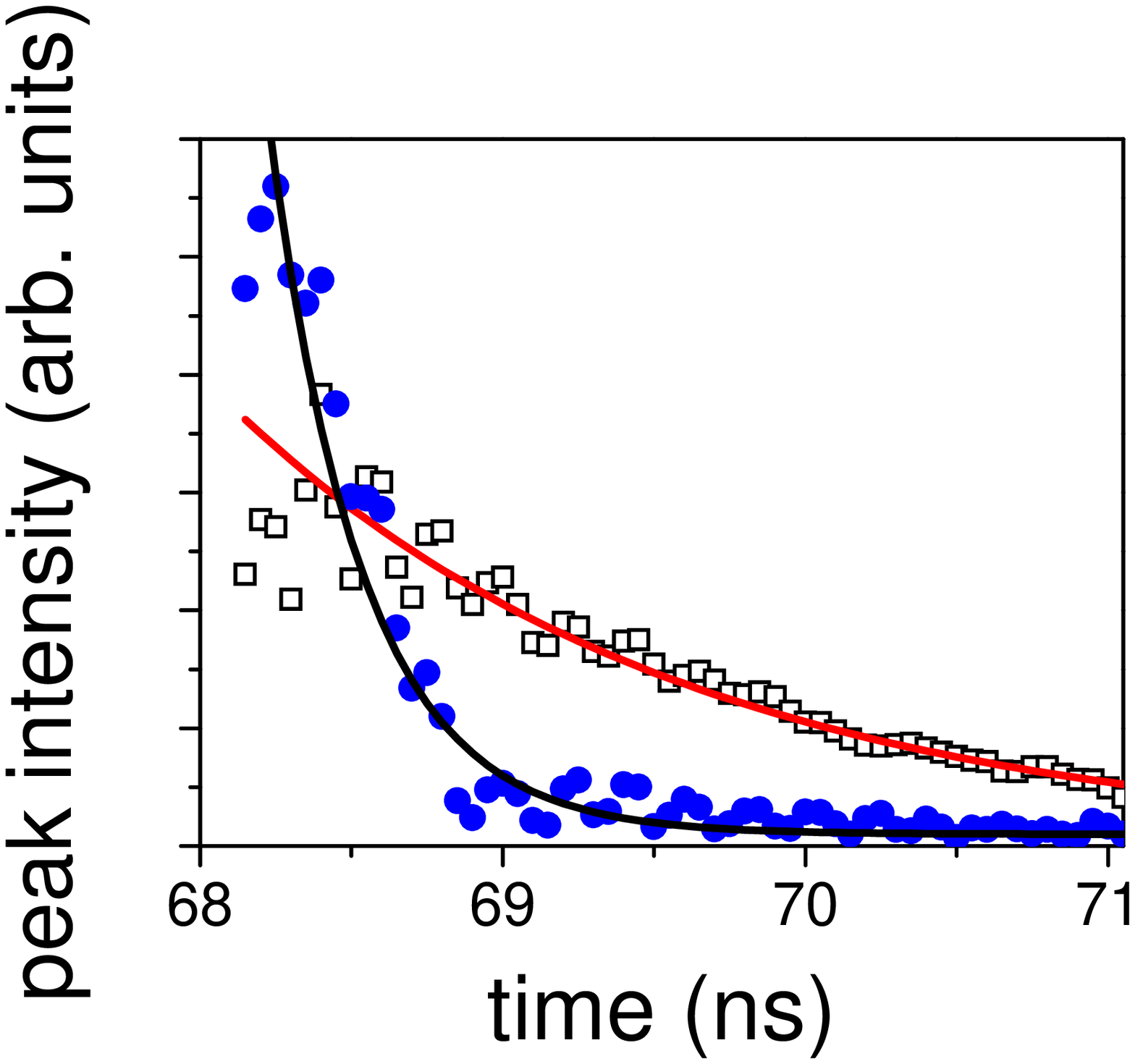} 
\caption{Configuration 1. Left: sliding Fourier transforms of the time derivative of Im($\Delta r/r$), for sample C1. The time indicates the center of the 300ps Fourier transform window. Right: peak intensity of the cavity mode (open symbols) and the lower band edge peak (closed symbols) as a function of time. The solid lines are exponential fits.}
\label{fig3}
\end{figure}
To analyse these results, we perform calculations of the acoustic field in the multilayers with a standard transfer matrix formalism\cite{perrinRossignol}. We calculate the electromagnetic field in each layer and assume that the interaction of the device with the incoming light leaves a stress proportional to the local electromagnetic field energy in GaAs layers only. The sound wave propagation equation is solved in each layer, taking into account this stress source term and boundary conditions to obtain the acoustic field. In particular we access to the 
deepest layer displacement (just before the substrate) which gives the acoustic field in the substrate. 
As we measure in $\Delta r/r$ 100GHz oscillations with an amplitude of $10^{-6}$, the strain incident on the aluminum film  can be estimated to $\eta \sim 6.10^{-7}$ ($dn/d\eta = -28+15i$ \cite{DijkhuisAlu}). At such strain, non linear effect are negligible.We then calculate the imaginary part of $\Delta r/r$, including contributions of the surface displacement and photoelastic effects in aluminium. Comparison with the experimental data is reported fig.~2b, and shows an excellent agreement. 
The same conclusion applies at higher frequencies: figure 2 shows also results for sample C2 which emits phonons at 200GHz.

The cavity mode width deduced for C1 from a 3ns time integration 
is $\Delta f=0.5$GHz to be compared with the theoretical
limit 0.32GHz given by the quality factor. Another way
to analyse the lifetime of the generated phonon modes
is to perform sliding Fourier transform over 300ps time
windows whose central value scans the whole time range.
Results are reported on fig.~3 for sample C1: the
cavity mode lives much longer than the acoustic mirror
resonances in agreement with the fact that this mode tunnels
slowly outside the cavity owing to the confinement.
We can deduce an exponential decay of  $1300\pm 200$ps for 
C1, in agreement with the measured linewidth $\sqrt{3}/(\pi \Delta f)$ (1100ps).

To summarize we have demonstrated in the first part of the letter that  SLs and cavities can efficiently generate monochromatic phonons which can escape from the device and propagate through a thick substrate before being detected on the other side via a metallic transducer. 

Let us now consider the detection performance of these nanostructures. 
It is well established that in a semi-transparent material, the interaction of light with coherent
phonons is strongly enhanced at a particular
acoustic wave vector $q=2k$ where $k$ is the electromagnetic
wave vector in the sample\cite{thomsenMaris}. The same modes are Raman active in backscattering configuration. At the experimental wavelength, q is close to the reduced Brillouin zone edge in samples M1 and C1, and in the middle of the Brillouin zone in samples M2 and C2. The Brillouin mode corresponds to the lowest frequency. This enhanced detection has been demonstrated in the second set of
experiments where the pump beam is absorbed in the
aluminum layer deposited on the substrate, generating a broadband
acoustic pulse due to the short optical absorption
length.  In this case, propagation through the whole substrate strongly modifies the spectrum of the pulse and high frequency conversion is expected (configuration 2, see inset
in fig. 4 where the strain generated by the aluminium film and the same strain after nonlinear propagation are schematically shown). The detection of the resulting acoustic pulse by the device on the other side allows to study the detection process without mixing with the generation mechanism, and to take advantage of non linear frequency conversion to test the detector sensitivity at high frequencies. 

\begin{figure}[t]
\begin{center}
\includegraphics[angle=-90,width=0.48\textwidth]{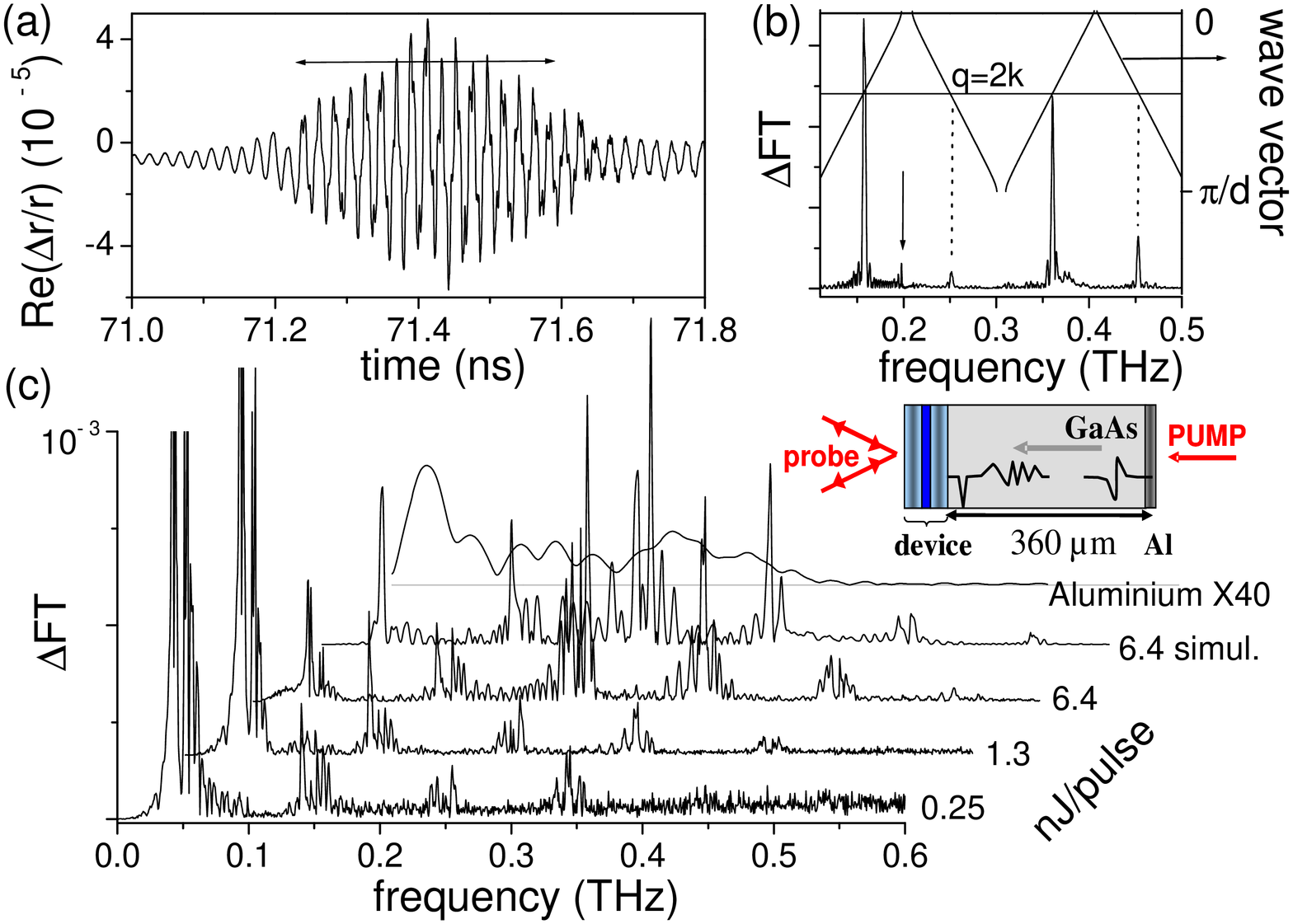} 
\caption{Inset: scheme of the experimental configuration 2. (a): Re($\Delta r/r$) as a function of the pump-probe delay on sample M2 (Arrows indicate the acoustic pulse round trip in the device). (b): Corresponding spectra, compared with the folded dispersion. (c): amplitudes of the time derivative of Re($\Delta r/r$), normalized to 1nJ/pulse, measured (i) on sample M1 at different pump intensities compared to a calculation for 6.4nJ/pulse, (ii) on a sample where the detector is an aluminum film at 7.6nJ/pulse.   }
\label{fig4}
\end{center}
\end{figure}
Figure 4a shows the real part of $\Delta r/r$ obtained on sample M2. The Brillouin oscillations can be observed on the whole time window whereas superimposed high frequency features occur during the round trip duration of the acoustic pulse in the multilayer, indicated by the arrows.  
Indeed, as previously said, the probe beam is not fully absorbed in the device and interacts in the substrate with the incoming acoustic pulse before it reaches the device, or when it leaves the device, after reflection on the free surface. 
Figure 4b shows the $\Delta $FT of this temporal trace. As expected, few discrete frequencies corresponding to the enhanced detection selection rule are observed among Bragg oscillations due to device finite size. Moreover, the detection is also amplified near the edges of the acoustic gaps where the density of state is enhanced by the modulation (see arrow around 200GHz). We show in figure~5 sliding $\Delta $FT calculated over 150ps time window for sample C1. As already mentioned, detection is enhanced close to the zone edge in this structure and the lowest folded mode almost coincides with the Brillouin oscillation. The Brillouin mode is detected first while folded modes are only observed during the time they remain in the device. Besides, we see the long-lived cavity modes within the zone boundary acoustic gaps (50, 150 GHz) and the zone center one at 100GHz (at 200GHz the gap is closed). This demonstrates that cavity modes can be excited just hitting the device with an acoustic pulse. They are clearly the only surviving frequencies after about 1000ps. In contrast to the mirror modes, these modes can be detected during the long time it takes them to leave the cavity. The longest time is obtained at 100GHz as C1 sample has been designed to optimize the acoustic gap width around this frequency.

\begin{figure}[tbp]
\includegraphics[angle=-90,width=0.4\textwidth]{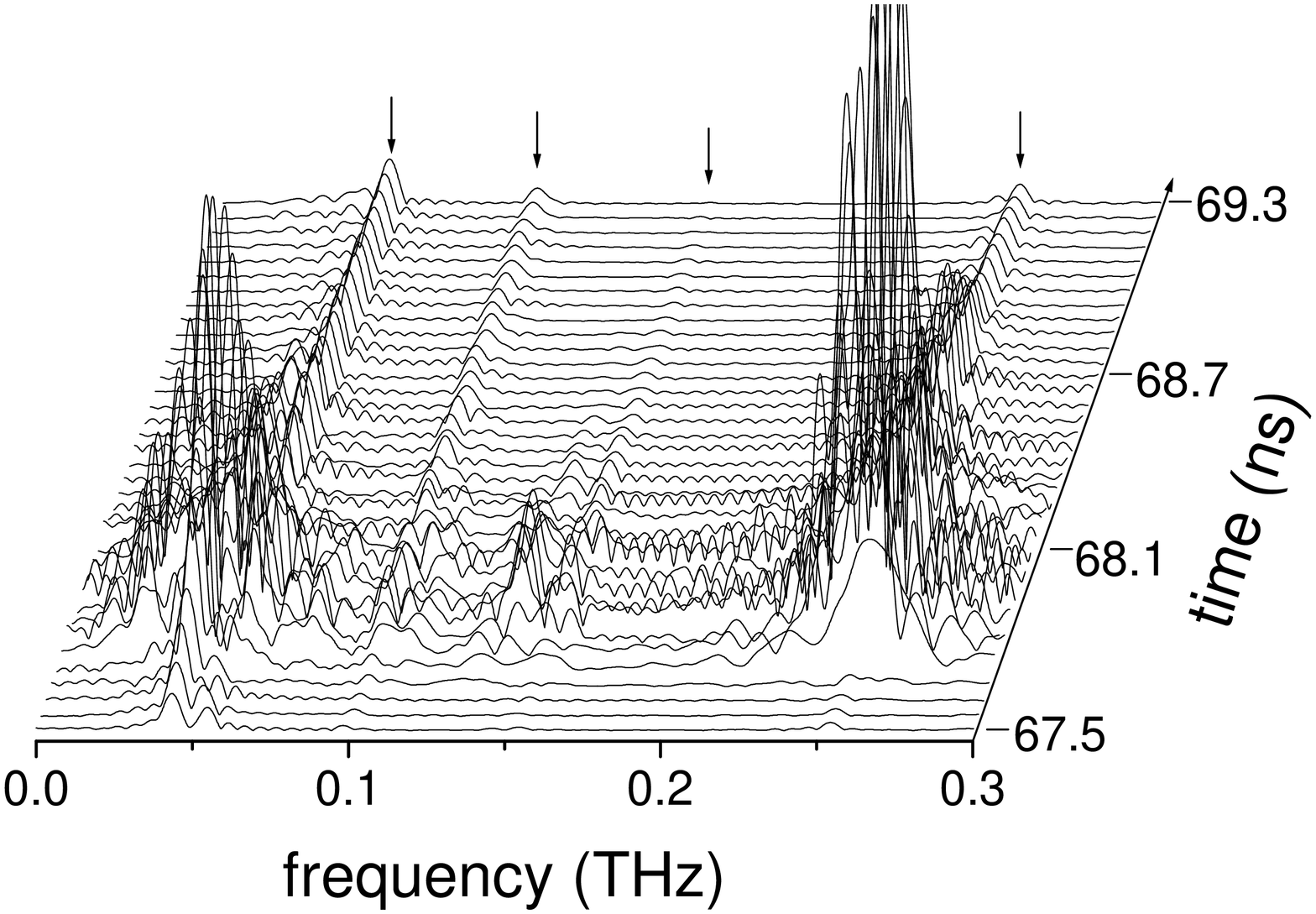} \vspace*{0cm}
\caption{Configuration 2. 200ps sliding Fourier transform of the time derivative of Re($\Delta r/r$) for sample C1. The time scale indicates the center of the Fourier transform window. Arrows show the cavity modes at 0.05, 0.15 and 0.25THz for zone boundaries and 0.1THz for the zone center. }
\label{fig5}
\end{figure}

Let us now consider the effect of high frequency transfer during the non linear propagation in the GaAs substrate.
Figure 4c shows the $\Delta $FT of Re($\Delta r/r$) normalized by the pump pulse energy measured in sample M1 with different energies from 0.2 to 6.4nJ. As in C1, the signal
is large around the zone boundary gaps. On the curve
obtained with a low intensity pump (0.25nJ/pulse), the
frequency range detected in M1 extends to 350GHz while for the largest intensity, frequencies up to 550GHz have been detected. We also display on fig 4c a simulation done for 6.4nJ/pulse pump intensity. Distorsion during propagation through the substrate due to sound dispersion, sound attenuation and non linear effects can be described by a 1D Korteweg-De Vries equation\cite{marissoliton,muskens} as far as diffraction is negligible. 
Sound absorption is negligible since we work at low temperature. Solving this equation, the spectrum of the acoustic pulse incident on the device has been calculated and Re($\Delta r/r)$ of the device is obtained with a standard method\cite{perrindetection1,perrinRossignol}. The incoming strain on the device can be estimated to $6.10^{-4}$. In similar experiments performed with an aluminum detector\cite{peronnesoliton}, 
i. e. on a GaAs substrate with aluminum on both sides, 
frequencies up to 200GHz have been observed with a low power excitation
and up to 350GHz for 7.6nJ/pulse as llustrated in fig.~4c. Moreover, the detected signal was more than one order of magnitude weaker. These results demonstrate that semiconductor multilayers are in comparaison very efficient detectors at high frequencies.

In summary, we have introduced a new experimental scheme combining a white acoustic transducer (aluminum film) and a quasi monochromatic one (semiconductor multilayer). This allowed us to perform two sets of experiments which demonstrate independently that phonon mirrors and acoustic cavities are efficient quasi-monochromatic phonon generators and also selective and sensitive phonon detectors. We show that standard models provide a reasonable understanding of our results, nevertheless, these experimental configurations will allow future detailed studies of generation and detection mechanisms.
As the aluminum film and the multilayer are deposited on the opposite sides of a thick substrate, in contrast with experiments performed in front configuration, we clearly show that coherent phonons generated in the device can be efficiently  transferred into the substrate where they can propagate over millimetric distances.  This transduction efficiency should work in the THz range provided that two such devices are used for the  generation and detection. This would allow phonon imaging and spectroscopy in nanostructures  using acoustic waves with nanometric wavelengths.


\end{document}